\documentclass[twocolumn,trackchanges]{aastex631}
\usepackage{cancel} 
\usepackage{soul}

\shorttitle{{\em Astrosat}-{\em NICER} view of 4U 1702-429}
\shortauthors{Chattopadhyay et al.}

\graphicspath{{./}{figures/}}

\begin{document}

\title{Spectro-temporal evolution of 4U 1702-429 using {\em AstroSat-NICER}}

\correspondingauthor{}
\email{E-mail:soma2778.wbes@gmail.com}

\author[0009-0004-7264-3184]{Suchismito Chattopadhyay}
\affiliation{Government Girls' General Degree College, 7, Mayurbhanj Road, Kolkata 700023, India}

\author[0000-0002-7609-2779]{Ranjeev Misra}
\affiliation{Inter University Center for Astronomy and Astrophysics, Ganeshkhind, Pune 411007, India}

\author[0000-0002-1047-9911]{Soma Mandal}
\affiliation{Government Girls' General Degree College, 7, Mayurbhanj Road, Kolkata 700023, India}

\author[0000-0002-7567-3475]{Akash Garg}
\affiliation{Inter University Center for Astronomy and Astrophysics, Ganeshkhind, Pune 411007, India}

\author{Sanjay K Pandey}
\affiliation{Department of Mathematics, Shri L B S Degree College, Gonda 271003, India}

\begin{abstract}

We present the broadband spectral and timing properties of the atoll source 4U 1702-429 using two observations of {\em AstroSat} with the second one having simultaneous {\em NICER} data. For both observations, the spectra can be represented by a Comptonizing medium with a black body seed photon source which can be identified with the surface of the neutron star. A disk emission along with a distant reflection is also required for both spectra. For the first observation, the coronal temperature ($\sim 7$ keV) is smaller than the second ($\sim 13$ keV), and the disk is truncated at a larger radius, $\sim 150$ km, compared to the second, $\sim 25$ km, for an assumed distance of 7 kpc. A kHz QPO at $\sim 800$ Hz is detected in the first and is absent in the second observation. Modeling the energy-dependent r.m.s and time lag of the kHz QPO  reveals a corona size of $\leq$ 30 km. A similar model can explain the energy dependence of the broadband noise at $\sim 10$ Hz for the second observation. The results suggest that kHz QPOs are associated with a compact corona surrounding the neutron star and may occur when the disk is truncated at large distances. We emphasize the need for more wide-band observations of the source to confirm these results.

\end{abstract}


\section{Introduction} \label{sec1}

\noindent
Neutron star low-mass X-ray binaries (NSLMXB) are well known for showing variabilities on different time scales. These variabilities are believed to originate mostly from the nearby surroundings of the neutron star as it accretes gaseous matter from a low-mass companion star via Roche lobe overflow \citep{klitzing:frank1985accretion}. The Fourier analysis of variabilities along with the observed photon spectrum, indicate the presence of a truncated accretion region around the neutron star. Specifically, an optically thick and geometrically thin accretion disk around the neutron star gets truncated at some radius \citep{klitzing:hanawa1989x, klitzing:1992MNRAS.256..545L, klitzing:Smale2001, klitzing:2002MNRAS.331..453D,  klitzing:2003A&A...405..237B,klitzing:2006A&A...460..233C, klitzing:2018ApJ...867...64C} and the inner region is occupied by a Comptonized coronal cloud which is typically hotter than the disk and the boundary layer of the neutron star \citep{1973A&A....24..337S,1977ApJ...218..247L}.

The outer accretion disk predominantly emits black body photons, constituting the soft component of the observed photon spectrum. The Comptonizing corona, where the relatively cold input seed photons either from the disk, or boundary layer or from both, undergo multiple inverse Comptonization, forming the hard component of the observed spectrum. Furthermore, these hard photons often illuminate the optically thick disk and create a reflected spectrum that consists of multiple emission lines and a Compton hump. Despite the inherent narrowness of these Fe K lines, they get broader under the influence of gravitational Redshift and Doppler shift \citep{klitzing:2000PASP..112.1145F, klitzing:ballantyne2001x}. Since the absorption cross section reduces with energy, high energy photons instead of getting absorbed above 10 keV get Compton down scattered and produce a hump-like feature in the spectra\citep{klitzing:ross2007x,klitzing:10.1093/mnras/stv304}.

Besides showing rich features in the energy spectra, NSLMXBs also show quasi-periodic oscillations (QPOs)  in their Power Density Spectra (PDS) \citep{klitzing:van1989quasi, klitzing:van1997kilohertz, klitzing:peille2015spectral}. Categorized by frequency, QPOs are generally grouped into three classes: Lower Hertz QPOs (5-60 Hz), Hecto Hertz QPOs (100-200 Hz), and Kilo Hertz QPOs (200-1300 Hz). These QPOs are characterized by three key attributes, namely centroid frequency (f), quality factor (Q), and normalization. The centroid frequency carries the information of the dynamical time scale of the corresponding system, whereas the quality factor indicates how coherent the oscillation is  \citep{klitzing:Wang2016ABR,2019NewAR..8501524I}. As QPOs are believed to originate from regions close to the neutron Star, they can provide vital insights into the physics of extreme gravity and how it affects the geometry of the inner accretion region. 

QPOs exhibit energy-dependent properties such as fractional root mean square and time lag which represent the delay between the high and low energy photons. The time-lags for the kHz  QPOs are of the order of $\sim$ 50 microseconds. However, while the lower kHz QPOs show soft lags \citep{klitzing:barret2013soft} that is, the soft energy photons are delayed compared to the harder ones, the upper kHz QPOs show hard lags, indicating perhaps a different origin of the two features although they occur in pairs \citep{klitzing:de2013time, klitzing:peille2015spectral, klitzing:10.1093/mnras/stab1905}. The short time-lags of $\sim$ 50 microseconds indicate that they should be associated with the Compton scattering time-scale in a medium of size $\sim$ 10 km \citep{klitzing:lee1998comptonization}. In general, one expects that Compton scattering should produce hard lags, but soft lags can be produced and such a model can explain the observations, if there is a fraction of the Comptonized photons impinging back in the soft photon source \citep{lee2001compton, klitzing:kumar2014energy, klitzing:kumar2016constraining,2020MNRAS.492.1399K,2022MNRAS.515.2099B}. On the other hand time-lags for lower frequency QPOs ($\sim$ 10 Hz), as well as broadband noise, are significantly larger ($\sim$ 10 milliseconds).

The LMXB 4U 1702-429, has been identified as an atoll source \citep{1976IAUC.3010....1S,1991A&A...250..389O}. The inclination of the disk is reported to lie between 53-64 degrees \citep{klitzing:2019ApJ...873...99L}, and the distance to the source has been estimated to be $\sim$ 5.6 kpc \citep{klitzing:2008ApJS..179..360G} from the analysis of the Type I burst observed in its light curve. Burst oscillations are also detected in the source at a frequency $\sim$ 330 Hz using {\it RXTE} observation \citep{klitzing:markwardt1999observation}. Using the FPS (F. Douchin-P. Haensel-Skyrme type interaction) model \citep{1994ApJ...424..823C} with an assumption of mass of the central compact object to be of 1.4$M_\odot$ and a typical radius of 10 km, the dimensionless spin parameter for this source has been obtained to be 0.155 \citep{klitzing:braje2000light}. Using {\em NuSTAR} \citep{klitzing:2019ApJ...873...99L}, {\em XMM-NEWTON} and {\em INTEGRAL} observations  \citep{klitzing:2016A&A...596A..21I}, it is revealed that the source has a high Fe abundance along with a higher disk ionization parameter ($\log \xi$) around 4. Moreover, kHz QPO around 722 Hz has also been detected using {\em RXTE} observations \citep{klitzing:markwardt1999observation}.

 There have been some recent studies on the spectral evolution of the 4U 1702-429. \citet{2024MNRAS.529.2234V} have studied the spectral evolution and characterized the  Type I thermo-nuclear bursts, using the same {\em AstroSat} observations of this source as analyzed in this work. Using the peak burst flux, they estimate the source distance to be $\sim 7$ kpc for non-isotropic emission and $\sim 8.5$ kpc for an an-isotropic one. The former distance estimate of $\sim 7$ kpc is somewhat higher than the one given by \citet{klitzing:2008ApJS..179..360G} ($\sim 5.5$ kpc) but is in the range reported from extinction measurement, $4.3-7$ kpc \citep{klitzing:2016A&A...596A..21I}. They fitted the persistent spectra using a power-law and a Comptonized disk emission. They found  power-law index $\sim 1.7$ and inferred an inner disk radius of $\sim 40$ and $20$ kms for the two observations. \citet{Banerjee_2024} have analyzed data from 14 {\em NICER} observations (which includes two observations used in this work) and one from {\em AstroSat} and {\em NuSTAR} to study the spectral evolution of the source.They found that the spectra can be described as thermal Comptonization of a black body emission along with a disk one. Spectral evolution of the source was identified with changes in the disk temperature and inner radius. They found evidence for an Iron line emission when the used {\em NuSTAR} and {\em AstroSat} data.

In this work, we study the broadband spectral and temporal properties of 4U 1702-429 using an {\it AstroSat} observation taken during 27/28 April 2018  and another simultaneous observation by {\it AstroSat} and {\it NICER} around 9 August 2019 \footnote{The NICER observations 2587030103 and 2587030101 are listed as Observations 9 and 10 in  \citet{Banerjee_2024}.}. The aim is to use the wide band capability of the instruments to constrain the energy spectra and then to quantify and model the temporal properties of the source.  Section \ref{sec2} describes the observations and the data reduction procedure, while Section \ref{sec3} deals with the light curves and hardness ratios.  The details of the spectral analysis have been discussed in Section \ref{sec4}. Section \ref{sec5} presents the timing analysis and the study of the energy-dependent temporal properties. Finally, the results of the work are summarized and discussed in Section \ref{sec6}. 

\section{Data Reduction} \label{sec2}
\begin{deluxetable*}{c c c c c c}
	\centering
	\tablecaption{Details of the observations of 4U 1702-429.\label{Table1}}
	\tablewidth{0pt}
	\tablehead{
		\colhead{Telescope} & \colhead{Obs. ID} & \colhead{Simultaneous} & \colhead{Obs. Date} &
		\colhead{Instrument} & \colhead{Exposure time} \\
	}
	
	\startdata
	{\it{AstroSat}} & A04\_225T01\_9000002062& No & 27-28 April, 2018 & { SXT}, {LAXPC } & 60 Ks\\
	{\it{AstroSat}} & A06\_002T03\_9000003080 & Yes & 8-10 August, 2019 & { SXT}, {LAXPC }  & 59 Ks\\
	{\it{NICER}} & 2587030101& Yes & 8 August, 2019 & {XTI } & 9.8 Ks \\
	{\it{NICER}} & 2587030102& Yes & 9 August, 2019 & {XTI } & 18.1 Ks\\
	{\it{NICER}} & 2587030103& Yes  &9 August, 2019 & {XTI } & 14.9 Ks\\
	\enddata
	
\end{deluxetable*}
{\em{AstroSat}} observed the source 4U 1702-429 on two occasions within a gap of almost a year, whereas {\em{NICER}} has observed it multiple times. The details of the observations are mentioned in Table \ref{Table1}.

We have used the software {\tt LAXPCsoftware22Aug15}, which is available on the official ASCC page \footnote{\url{http://astrosat-ssc.iucaa.in/laxpcData}} to analyze the LAXPC data. For each observation, we have merged all the orbit files to create the level2.event.fits files from the level 1 data. Further, scientific products have been generated using this merged level2.event.fits file and the good time intervals (gti) list, which removes the earth occultation and SAA. We have generated light curves, hardness ratio diagrams (HID), PDS, time lags, fractional rms variation, and energy spectra using LAXPC 20 as the LAXPC 30 unit had been switched off due to gain-related issues, and LAXPC 10  has a higher background count as one of its veto anodes is not functioning properly \citep{klitzing:2017ApJS..231...10A}. We have followed all the standard procedures as documented in ASCC page \footnote{\url{http://astrosat-ssc.iucaa.in/uploads/threadsPageNew\_SXT.html}}.

{\tt SXTPIPELINE}\footnote{\url{https://www.tifr.res.in/~astrosat\_sxt/sxtpipeline.html}} has been used to process the SXT level1 data to produce the level2 data. Level2 data from different orbits have been merged using {\tt SXTEVTMERGERTOOL} (Julia-based module)\footnote {\url{http://astrosat-ssc.iucaa.in/sxtData}} to produce the merged event file, which is further analyzed using Heasoft v6.29 routine {\tt XSELECT}. The count rates for both of the observations are well below 40 counts/sec, and thereby pile-up correction is not needed for the SXT analysis. We have chosen a region of 720 arcsec to produce the SXT spectra and light curves.

\begin{figure*}
	\centering
\includegraphics[scale=0.5]{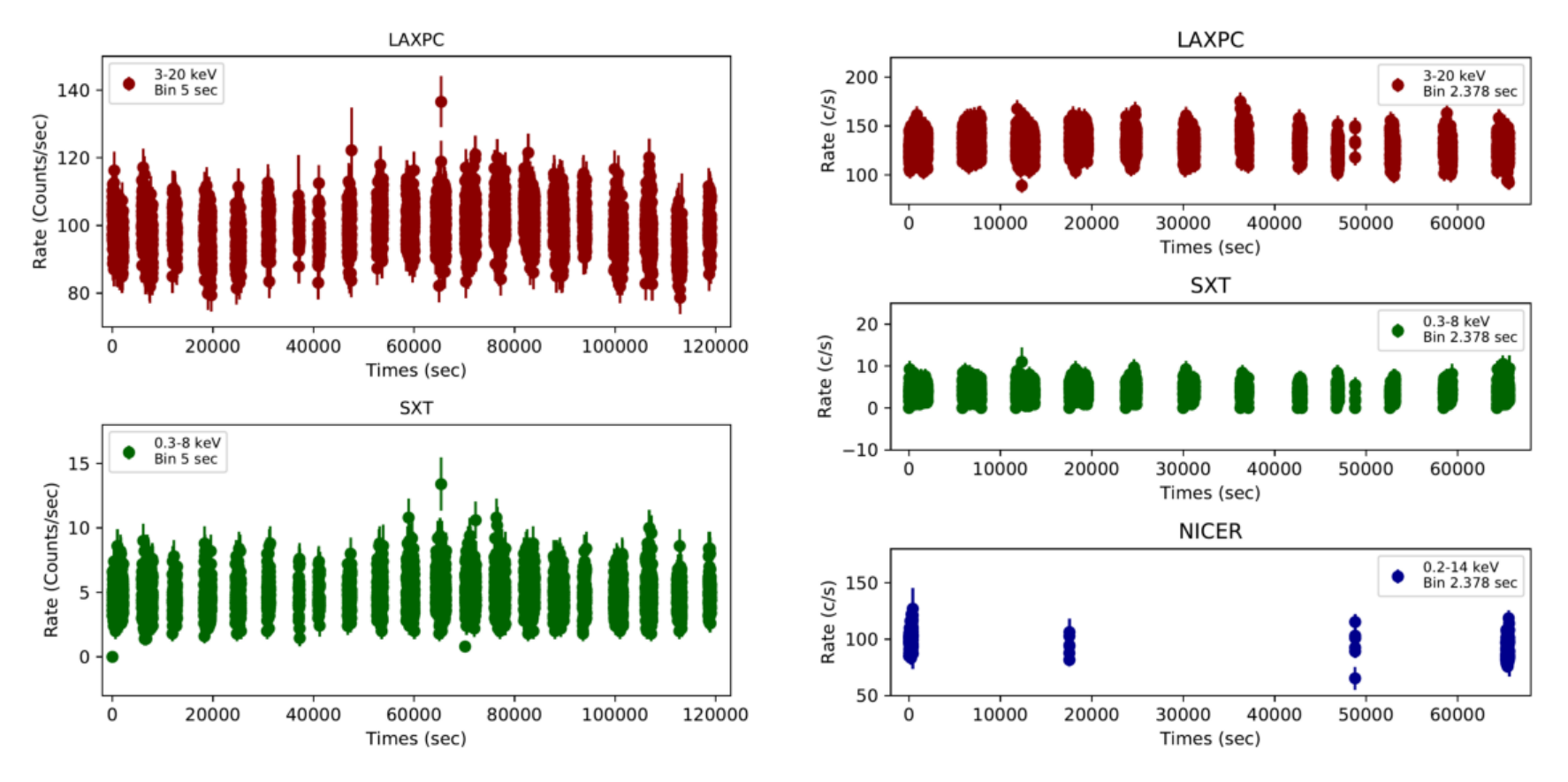}
	\caption{The left panel shows the simultaneous background-subtracted LAXPC 20 (Top) and SXT light curve (Bottom) for Observation 1 of 4U 1702-429 in the energy range of 3-20 keV and 0.3-8 keV, respectively. The right panel shows the simultaneous LAXPC 20 (Top), SXT (Middle), and  {\em NICER} (Bottom) light curve for Observation 2 using energy band of 3-20 keV, 0.3-8 keV, and 0.2-12 keV, from top to bottom panel respectively. All the light curves have the same binning time of 2.378 seconds.}
	\label{Fig1}
\end{figure*}

We have used {\tt nicerl2} pipeline\footnote{ \url{https://heasarc.gsfc.nasa.gov/lheasoft/ftools/headas/nicerl2.html}} for analyzing the {\em NICER} data. The final clean event file has been generated by omitting the effects of the detector 14 and 34 using Heasoft v6.29 routine {\tt FSELECT} as the increment in the noise was detected by the team in both detectors\footnote {\url{https://heasarc.gsfc.nasa.gov/docs/nicer/data\_analysis/nicer\_analysis\_tips.html}}. The   {\em NICER} CALDB (version 20221001) has been used during the extraction process. For the estimation of the background, {\tt nibackgen3c50} with gain epoch 2019 \footnote{ \url{https://heasarc.gsfc.nasa.gov/docs/nicer/analysis\_threads/background/}} has been used. We have merged the three {\tt NICER} clean event files and good time intervals list using {\tt nimaketime} and generated the science products using {\tt XSELECT}. 

\begin{figure}
	\includegraphics[width=1.0\linewidth]{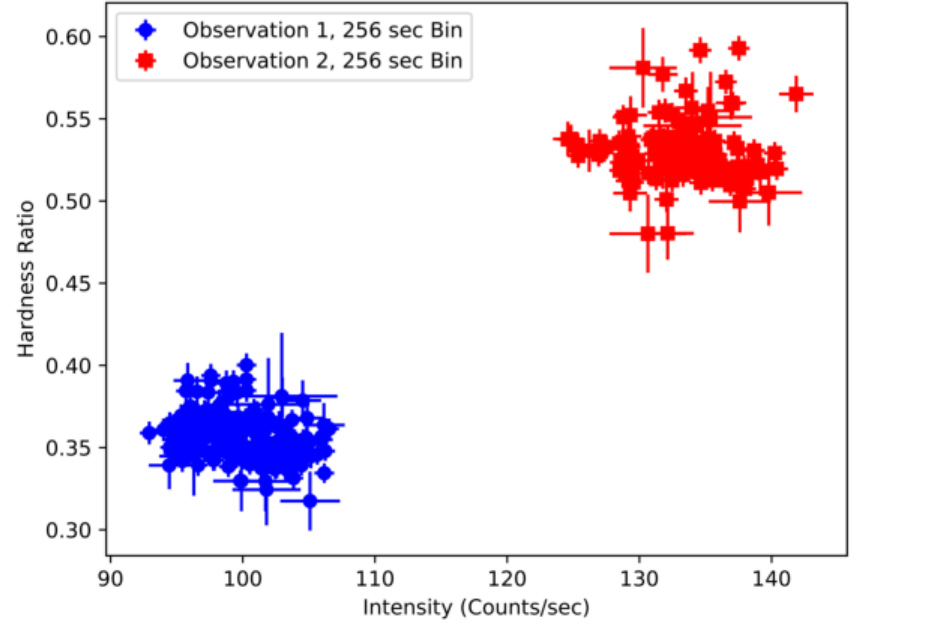}\\
	\includegraphics[width=1.0\linewidth]{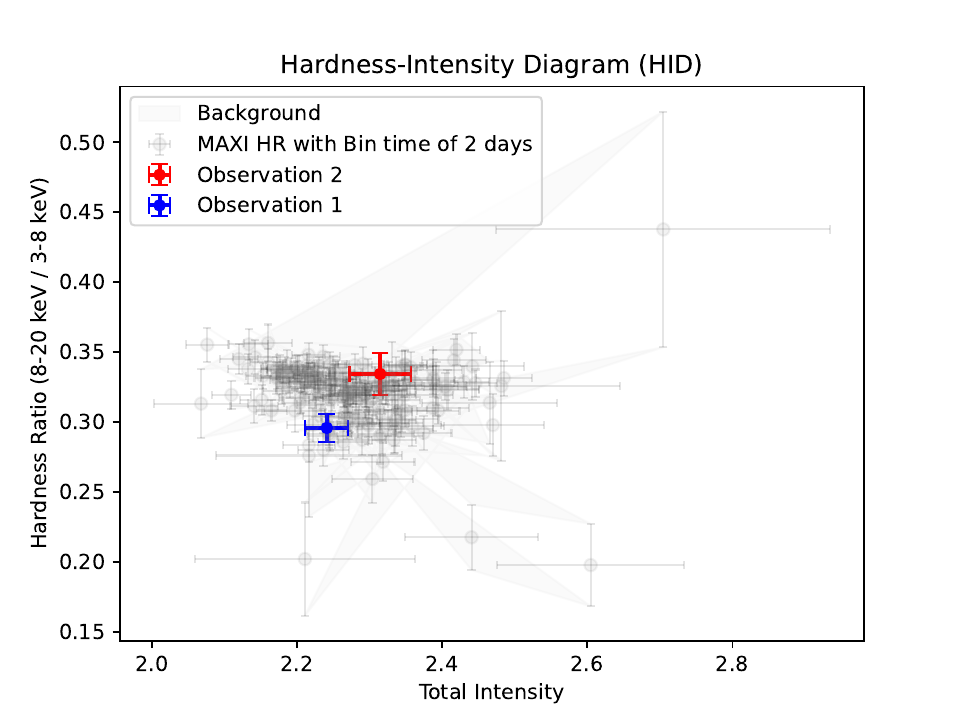}
	\caption{Top panel shows Hardness Intensity Diagram for the Obs $1 \& 2$ using background subtracted light curves of 3-8 keV (soft band) and 8-20 keV(Hard band) with time bin 256 seconds using LAXPC 20 only whereas bottom panel shows the position of Obs $1 \& 2$ in comparison to the MAXI hardness intensity diagram with a time bin of 2 days. The intensity is defined to be the sum of the count rates in the given 2 energy bands and the hardness ratio is defined to be the ratio of the counts in hard band with respect to the soft band..\label{Fig2}}
\end{figure}


Furthermore, the light curves from all the observations in SXT, LAXPC, as well as {\em NICER}, showed clear evidence of the presence of Type I bursts, which has been removed manually by editing the good time intervals. The joint SXT, LAXPC, {\em NICER}  energy spectra are analyzed using the Heasoft routine {\tt XSPEC} (version 12.12.0)\citep{klitzing:1996ASPC..101...17A}.

\section{Light curves and HIDs} \label{sec3}
The left panel of Figure \ref{Fig1} displays the 2.378-second (time resolution of SXT) binned background subtracted and simultaneous light curve of SXT and LAXPC for the observation ID A04\_225T01\_9000002062 (hereafter, Obs. 1). Top and bottom panels show the LAXPC 20  and SXT light curve in the energy range of 3-20 keV and 0.3-8 keV respectively. It is to be noted that there were two Type I bursts in the light curve which have been removed manually since the focus here is to study only the non-burst behavior of the source.

The right panel of Figure \ref{Fig1} shows the LAXPC 20 and SXT light curve for observation ID A06\_002T03\_9000003080 (hereafter, Obs. 2) with the same time bin of 2.378 sec. As the source does not show much variation in the HID, we have merged all the {\em NICER} observations and generated the 2.378 sec binned light curve in 0.2-12 keV as shown in the bottom panel of the right column of the Figure \ref{Fig1}. Here also, we have removed the single Type I burst from the light curve of SXT, LAXPC and two Type I bursts from the merged {\em NICER} light curve. All the light curves are barycenter corrected. The energy spectrum has been studied after taking into account these strictly simultaneous regions of the three observations.

The top panel of Figure \ref{Fig2} shows the variation of hardness-ratio (ratio of the counts of the hard band (8-20 keV) to softer one (3-8 keV)) with intensity for Obs. 1 $\&$2 using LAXPC 20. We find that the source is in a softer state in Obs. 1 in comparison to Obs. 2. As there are only minor variations in both intensity and hardness ratio, we have chosen not to segment further the HID for each observation. The bottom panel of Figure \ref{Fig2} illustrates the long-term evolution of the source in the hardness ratio diagram using a year-long MAXI light curve. Each single point in the bottom panel of Figure \ref{Fig2} represents the source's position in every two days. 

\section{Spectral Analysis}  \label{sec4}
We have performed joint spectral analysis using simultaneous data from LAXPC, SXT and {\em NICER} XTI in a broad energy range of 0.7-20 keV. We consider the LAXPC spectrum in 4-20 keV only as the background count rate begins to dominate beyond 20 keV, while the SXT spectrum lies between 0.7-7 keV. Further, we take {\em NICER} spectrum in 0.7-10 keV energy range as systematic beyond 10 keV increases sharply up to 40\% from 1.5\% at lower energies\footnote {\url{https://heasarc.gsfc.nasa.gov/docs/nicer/analysis\_threads/spectrum-systematic-error/}} whereas the count rate is low below 0.7 keV. All the spectra are optimally grouped using the {\tt{ftgrouppha}} and a systematic of 2\% for the {\em NICER} and 3\% for joint SXT and LAXPC spectra have been used. Additionally, a gain correction with a slope fixed at 1 has also been added for the SXT spectra only throughout the analysis. 

\begin{figure}
	\includegraphics[width=1.0\linewidth]{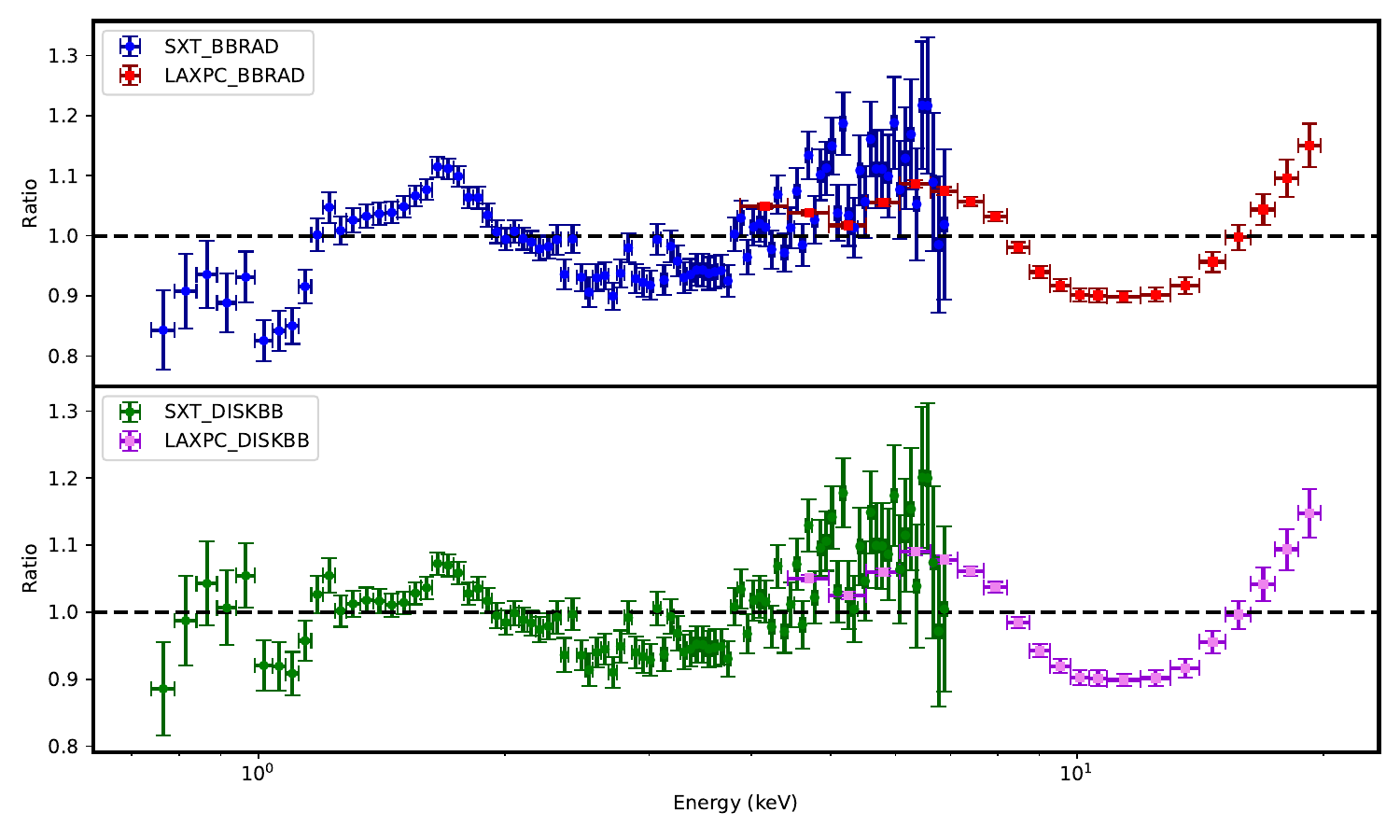}
	\caption{The upper panel shows the ratios of the spectral fitting using {\tt cons*tbabs*(thcomp*bbodyrad)}. High excess residuals can be seen at 2 keV, between 5-10 keV, and beyond 12 keV. The bottom panel shows the ratios of the spectral fitting using {\tt cons*tbabs*(thcomp*diskbb)}. The excess residuals are still visible.\label{Fig3}}
\end{figure}

\subsection{Observation 1}

We start by modeling the spectrum as an absorbed Comptonized emission produced by inverse Comptonization of the input soft photons from the boundary layer of the neutron star. Thus, we have fitted the joint LAXPC and SXT spectra using the combination {\tt constant*(tbabs*(thcomp*bbodyrad))} where the constant takes care of the cross-calibration between the LAXPC and SXT spectra. We describe the thermal Comptonization of input black body photons using the convolution model {\tt thcomp} \citep{klitzing:2020MNRAS.492.5234Z} and {\tt  bbodyrad}. We use {\tt tbabs} \citep{klitzing:2000ApJ...542..914W} model to consider the Galactic absorption. As {\tt thcomp} is a convolution model, we have extended the energy range for model computation from 0.01 to 1000 keV with 500 logarithmic bins to fit the spectra in the 0.7-20 keV range. The reduced chi-square ($\chi_{red}^2$) for this fit came out to be greater than 2. We find significant residuals around 2 keV and between 5-10 keV, as shown in Figure \ref{Fig3}. Even if we consider {\tt diskbb} \citep{klitzing:1984PASJ...36..741M} as a seed photon source, we can't account for the residuals, indicating perhaps the presence of reflection features in the spectrum.

Next, we add a Gaussian component to the model combination to resolve the excess residuals. We have fixed the {\tt Gaussian} line energy at 6.40 keV during the spectral fitting and allow the FWHM (Full Width Half Maxima) and the norm to vary. This gives an improved $\chi_{red}^2$ of 109.95/105 and 110.49/105 for {\tt bbrad} and {\tt diskbb} respectively.  However, the normalization of the disk  or black body component came out too large, resulting in un-physically large radius of $> 300$ km  for an assumed distance of 7 kpc \citep{2024MNRAS.529.2234V}. Moreover, the width of the line was found to be $\sim 1.8$ keV, which is too broad to be explained by relativistic broadening from a disk having such a large inner radius. Hence, we considered other possibilities to model the residuals.

\begin{deluxetable}{c  c  c}
	\centering
	\caption{\label{Table2} The best-fitted value using the model {\tt const*tbabs*(Gaussian+ diskbb+ireflect*thcomp*bbodyrad)} for Obs. 1 and for Obs. 2. $n_H$ is the hydrogen absorption column density which is measured in the unit of $10^{22}$ atoms cm$^{-2}$. $kT_e$ is the Coronal electron temperature, $kT_{bb}$ is the black body temperature $\tau$ is the optical depth. $\xi$ is the ionization parameter and $^\dagger$ denotes that the parameter is kept frozen during the spectral fitting. Further, the inclination angle is held fixed at 55 degree, the covering fraction has been fixed  at 1, the disk ionization parameter $\xi$  is fixed at 4999 disk temperature associated with {\tt ireflect} is fixed at 30000 kelvin, Gaussian line energy at 6.7 keV and Gaussian width at 0.11 keV.  The flux has been reported with in the 0.7-30 keV using the routine {\it cflux}.}
	\tablewidth{0pt}
	\tablehead{
		\colhead{\bf Param} & \colhead{\bf (Obs 1)} & \colhead{\bf (Obs 2 \& NICER)}\\
	}
	\startdata
	Const  & 1/0.93 & 1/0.94/0.85 \\ 
	$n_H$  & $1.72^{+0.04}_{-0.04}$ & $1.40^{+0.06}_{-0.07}$ \\
	Gauss Norm $\times 10^{-4}$ & $1.31^{+3.01}_{-1.31}$ & $1.41^{+1.5}_{-1.41}$\\
	$kT_{in}$ (keV)& $0.41^{+0.02}_{-0.01}$ & $0.54^{+0.06}_{-0.06}$\\
	$N_{dbb}$ & $1894^{+710}_{-509}$& $131.7^{+99}_{-60}$\\
	rel\_refl  &  $0.51^{+0.23}_{-0.16}$ & $0.58^{+0.12}_{-0.16}$ \\
	Tau ($\tau$) & $5.36^{+0.05}_{-0.05}$ & $4.09^{+0.75}_{-0.91}$\\
	$kT_{e}$ (keV) &  $6.93^{+1.38}_{-0.92}$ &  $12.7^{+3.22}_{-4.18}$ \\
	$kT_{bb}$ (keV) &  $0.71^{+0.01}_{-0.01}$ & $0.72^{+0.01}_{-0.03}$  \\
	Blackbody\_norm  &  $211^{\dagger}$ &  $211^{\dagger}$  \\
 	Flux $\times 10^{-9}$ (in cgs) & $ 1.76^{+0.09}_{-0.09} $ & $1.82^{+0.1}_{-0.1}$ \\
    Luminosity$\times10^{36}$ (in cgs) & $3.69^{+0.05}_{-0.05}$ &  $8.80^{+0.18}_{-0.18}$\\
	$\chi^2$/Dof & 83.37/104 & 207.03/204 \\
	\enddata
	
\end{deluxetable}

\begin{figure*}
	\includegraphics[width=1.0\linewidth]{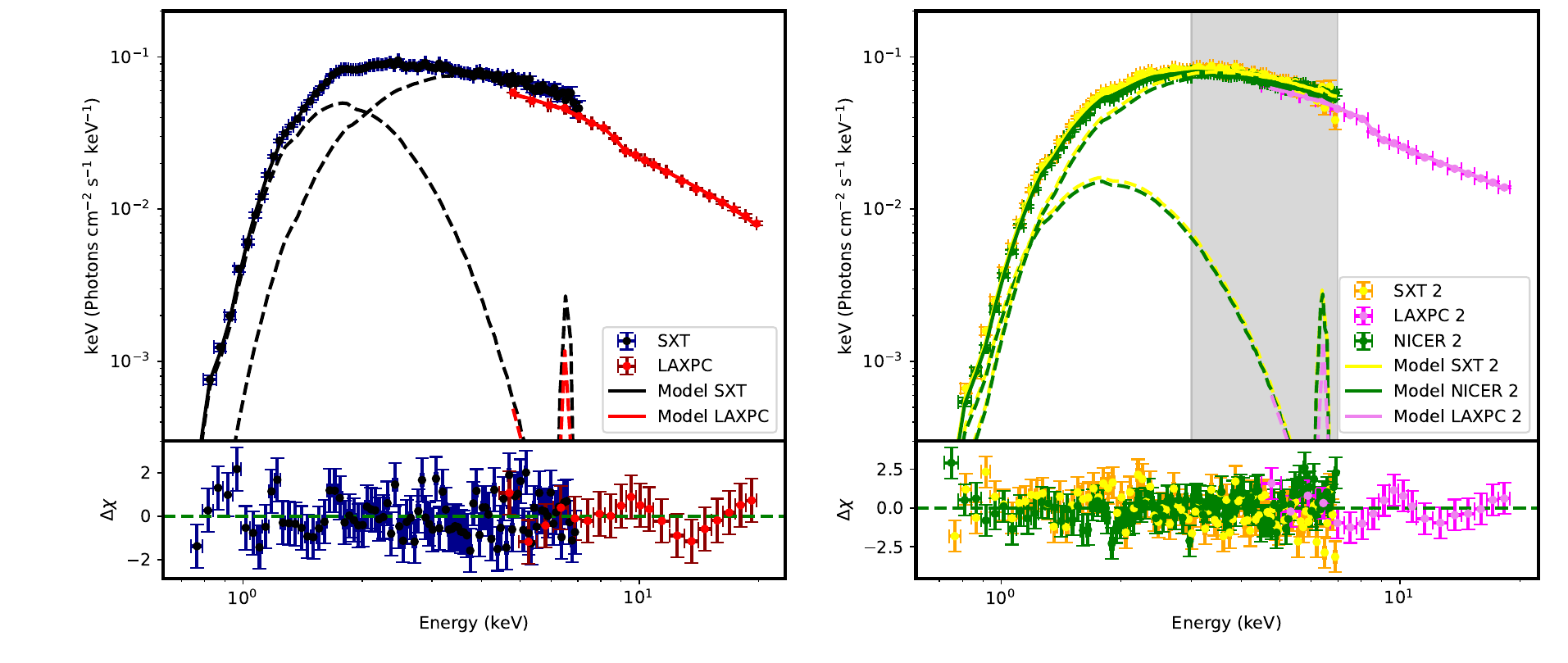}
	\caption{The left and right figures show the spectra for Obs. 1 and Obs. 2 respectively. Both the spectra have been fitted using the model {\tt cons*tbabs*(Gaussian+diskbb+ireflect*thcomp*bbodyrad)} and the bottom panels show $\Delta \chi$.}
	\label{Fig4}
\end{figure*}

We use the angle-averaged ionized reflection routine {\tt{ireflect}}, which models the emission due to the reflection of the Compton spectrum by the cold disk \citep{klitzing:1995MNRAS.273..837M}. This model allows the following free parameters: the reflection scaling factor $rel_{refl}$, the iron abundance $A_{Fe}$, and the disk ionization parameter $\xi$. Thus, we used the model combination {\tt const*tbabs*(ireflect*thcomp*bbodyrad)}.  
We have fixed relative Fe abundance to 1 and the disk inclination angle to 55 degrees based on earlier reports \citep{klitzing:2019ApJ...873...99L} while allowing the reflection fraction to vary. The $\chi^2/d.o.f$ comes out to be 116.9/106, which is an improvement over the previous fits. However, the  normalization{\footnote{https://heasarc.gsfc.nasa.gov/xanadu/xspec\\/manual/node137.html}} of the black body comes out to be significantly large 4$\times 10^5$. For a source distance of 7 kpc \citep{klitzing:2016A&A...596A..21I,2024MNRAS.529.2234V}, this corresponds to a radius of $\sim$ 150 km, much larger than the expected neutron star radius of 10 km. If, we fix the normalization to 211 (corresponding to a radius of 10 km ) the  $\chi^2/d.o.f$ is large 626.4/107.   Hence, we considered an additional soft thermal disk component {\tt diskbb} to take care of the excess soft energy residuals still present in the spectrum. Therefore, the model combination becomes {\tt const*tbabs*(diskbb + ireflect*thcomp*bbodyrad)}. Adding an extra soft component improved the reduced chi-square significantly (83.87/105). Since {\tt ireflect} does not include the iron line emission, we added a narrow {\tt Gaussian } at 6.7 keV, keeping the width fixed at 0.1 keV. This did not change the $\chi_{red}^2$ much resulting in 83.37/104. The best-fitted values are tabulated in Table \ref{Table2}. During the fitting, the best fit $\xi$  turned out to be the maximum allowed value  4999 ergs/$cm^2$ and hence was fixed at that value. Such a large value of disk ionization parameter $\xi$ has been previously reported for  \citep{klitzing:2019ApJ...873...99L} for this source. The {\tt diskbb} normalization of $\sim$ 1894 and a color correction factor \citep{1995ApJ...445..780S,1998PASJ...50..667K} of 1.7 yields an inner disk radius of $\sim$ $124^{+22}_{-18}$ km.

\subsection{Observation 2}
\noindent
We employed the same model combination of {\tt const * tbabs (Gaussian + diskbb + ireflect * thcomp * bbodyrad)}, with the same assumptions to Obs. 2 consisting of {\em AstroSat} + {\em NICER} spectra. We have obtained a satisfactory fit with a ${\chi^2/d.o.f}$ of 207.03/204. The best-fit parameters can be found in Table \ref{Table2} and the spectra are shown in the right of Figure \ref{Fig4}.

A significant difference between the spectra of Obs. 2 compared to 1, is that the temperature of the corona has increased from $\sim 7$ to $\sim 13$ keV, which explains the higher hardness ratio (in the LAXPC band) of Obs 2. The disk normalization has decreased implying an inner disk radius of $32.8^{+6.4}_{-15.6}$ km compared to $124^{+22}_{-18}$ km obtained for Obs 1. There is also a decrease in the absorption column density and an increase in the inner disk temperature.  

\begin{deluxetable}{c c c c}
	\centering
	\tablecaption{Details of the best-fitted parameters of the PDS for Obs. 1 and Obs. 2 using multiple Lorentzian components. $^{\dagger}$ denotes frozen parameters.\label{Table3}}
	\tablewidth{0pt}
	\tablehead{
		\colhead{\bf Parameters} & \colhead{\bf Obs.1 } & \colhead{\bf Obs. 2 }  &\colhead{\bf \em NICER} \\
	}
	\startdata 
	Frequency (Hz) &$0^{\dagger}$ & $0^{\dagger}$ & $0^{\dagger}$\\
	FWHM (Hz) & $44.54^{+7.40}_{-6.21}$ & $3.72^{+0.32}_{-0.33}$ & $3.89^{+0.35}_{-0.36}$\\
	Norm$\times 10^{-2}$ & $2.74^{+0.2}_{-0.3}$&$0.96^{+0.10}_{-0.10}$ & $1.68^{+0.07}_{-0.09}$\\
	Frequency (Hz) &$0^{\dagger}$ &$8.84^{+0.91}_{-1.2}$& $9.17^{+1.35}_{-2.39}$\\
	FWHM (Hz) & $> 479.18$ &$14.26^{+3.07}_{-2.60}$& $14.10^{+6.14}_{-3.95}$\\
	Norm$\times 10^{-2}$ & $6.96^{+1.80}_{-3.27}$&$0.95^{+0.02}_{-0.02}$&$1.02^{+0.10}_{-0.10}$\\
	Frequency (Hz) & $790.71^{+19.32}_{-19.22}$& - & -\\
	FWHM (Hz) & $140.33^{+74.40}_{-51.22}$& - & -\\
	Norm$\times 10^{-2}$ & $2.40^{+0.80}_{-0.80}$& - & -\\
	$\chi^2$/dof & 117.58/115 &60.72/70 & 48.70/47\\
	\hline
	\enddata
	
\end{deluxetable}

\section{Timing Analysis} \label{sec5}
\subsection{PDS}

\begin{figure}
	\includegraphics[width=1.02\linewidth]{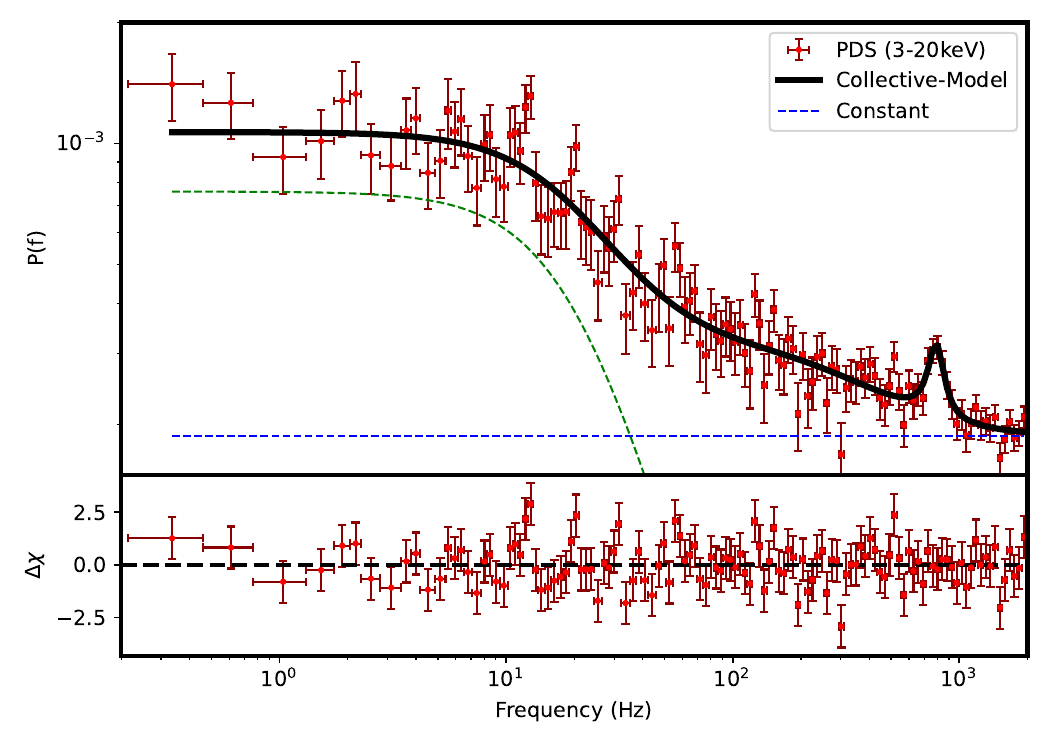}	\\
	\includegraphics[width=1.03\linewidth]{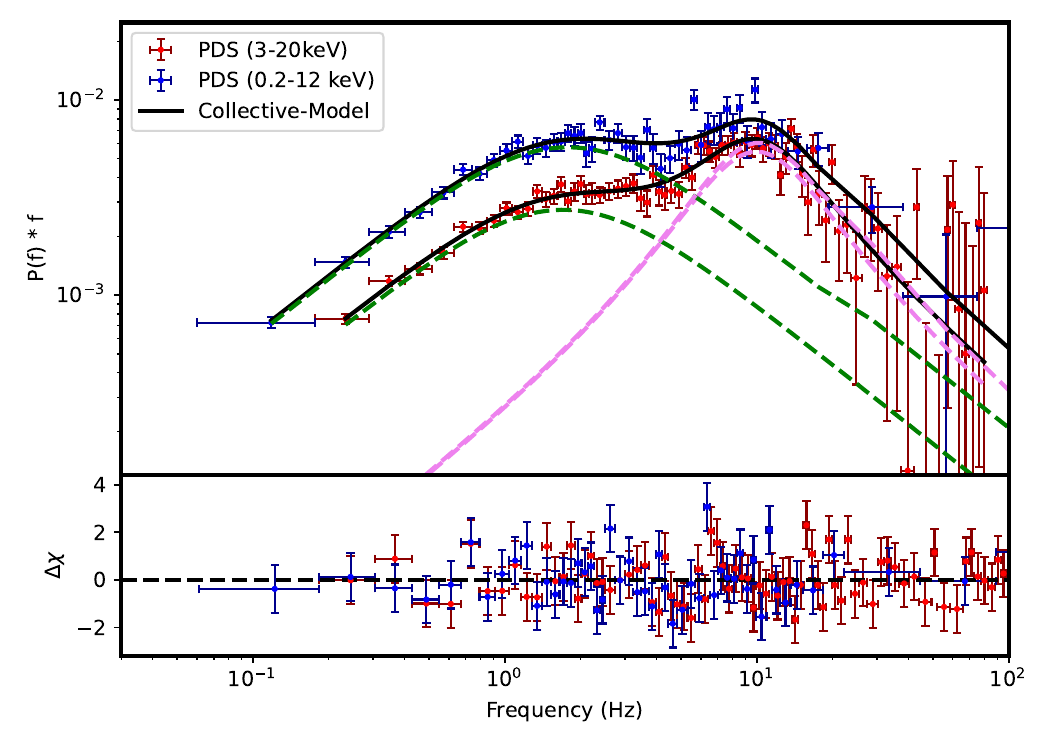}
	\caption{The best fitted PDS for Obs. 1, Obs. 2 and {\em NICER}. Blue points represent the {\em NICER} PDS and red points represent the LAXPC PDS.}
	\label{Fig6}
\end{figure}

We have generated Power density spectra (PDS) for LAXPC 20/{\em AstroSat} Obs.1 and Obs.2 using the standard routine {\tt laxpc\_find\_freqlag} in the energy band of 3-20 keV with a Nyquist frequency of 2000 Hz. Further, the PDS has been rebinned using the subroutine {\tt laxpc\_rebin\_power powfiles} considering the Signal to Noise Ratio (SNR) 5. The software subtracts the dead time (42 $\mu s$) corrected Poisson noise from the PDS \citep{2016ApJ...833...27Y}. However, there is faint constant component in the PDS, which we attribute to an underestimation of the Poisson level. Thus, we include a constant while fitting the PDS to take this into account. As shown in the top panel of Figure \ref{Fig6}, Obs. 1  shows a clear presence of a high-frequency QPO around $\sim$ 790 Hz with a quality factor \footnote{The quality factor is defined as the ratio of\\ the centroid frequency to the width of the QPO.}of $\sim$ 6 along with broad low-frequency noise.  \citet{2024MNRAS.529.2234V} have reported that between 0.1-100 Hz they did not find any QPO features in the LAXPC data, which is consistent with our results, that only a kHz QPO is detected in Obs 1. On the other hand, the PDS for the second observation does not show any significant kHz QPO and the bottom panel of Figure \ref{Fig6} shows the broadband noise in the PDS. Note that the frequency range for the bottom panel is different for the bottom panel and the frequency times power is plotted to illustrate more clearly the broadband noise seen in Obs. 2.

Further, we have generated the {\em NICER} PDS in the 0.3-12 keV energy band. We have used a 5 millisecond binned light curve to obtain the Nyquist frequency of 100 Hz for the PDS. We have created the averaged PDS using 27982 segments (each of length 81.92 sec with 16384 bins per segment) of the light curve and binned the PDS to obtain a (Signal to Noise Ratio) SNR $>$ 5, which is shown in the bottom panel of Figure \ref{Fig6}.

The LAXPC PDS for Obs 1 has been fitted using a Lorentzian for the kHz QPO and two zero-centered Lorentzian for the broadband noise. The LAXPC and {\em NICER} PDS for Obs 2 were fitted using two broad Lorentzian and the best fit parameters are listed in Table \ref{Table3}. Note that, as expected, the centroid frequency and the FWHM width for the two components for the {\em NICER} and LAXPC PDS are consistent with each other and the components differ by normalization.

\begin{figure}
	\includegraphics[width=1.01\linewidth]{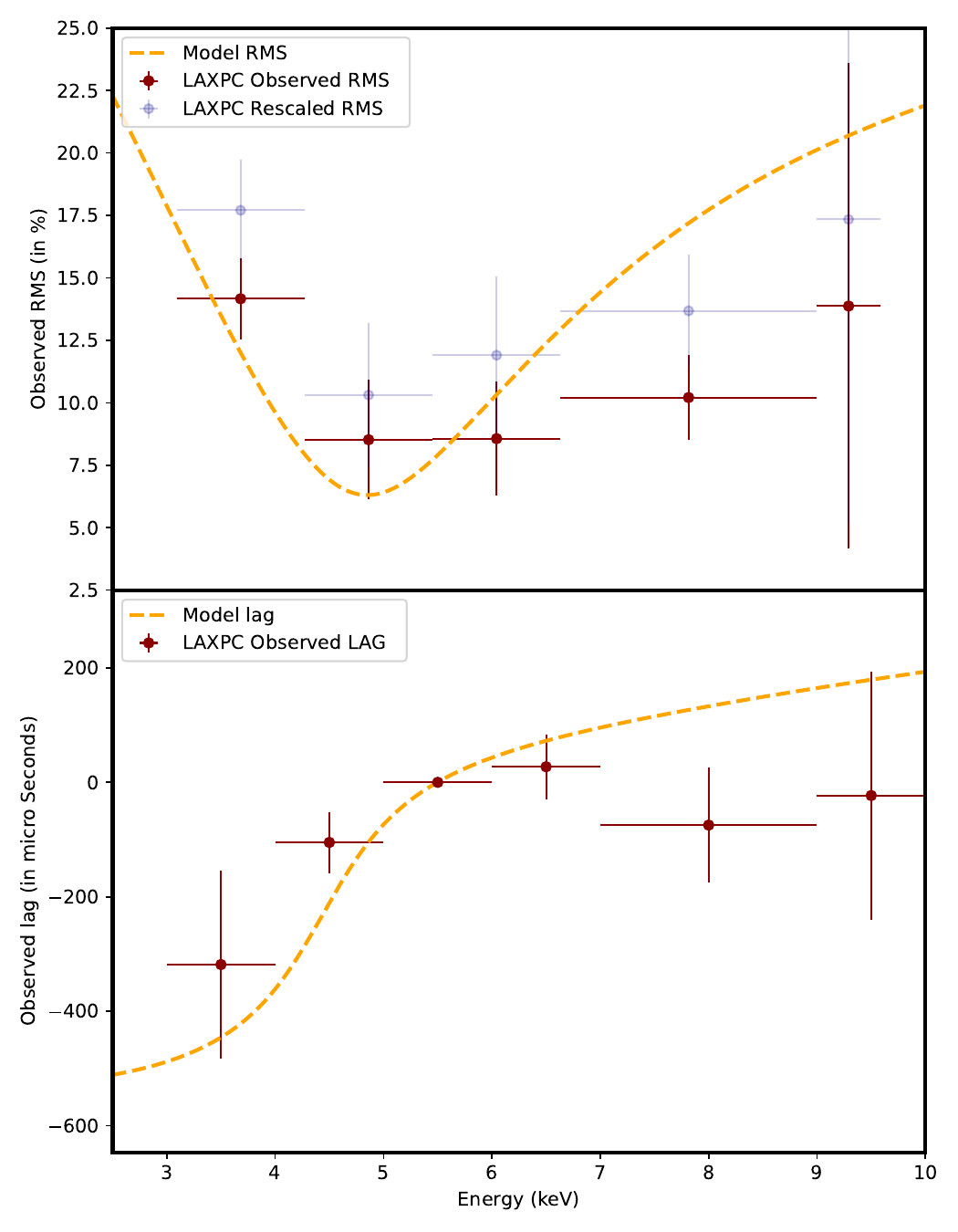}
	\caption{The Observed rms(top) and Lag(bottom) for the kHz QPO $\sim$ 800 Hz. The model lag and rms variation uses the variation in the heating rate of the corona. The light blue points indicate the re-scaled rms variation. The solid line represents the model prediction for a corona of size 30 km and $\eta$ = 0.1. }
	\label{Fig7}
\end{figure}

\begin{figure}
	\includegraphics[width=1.0\linewidth]{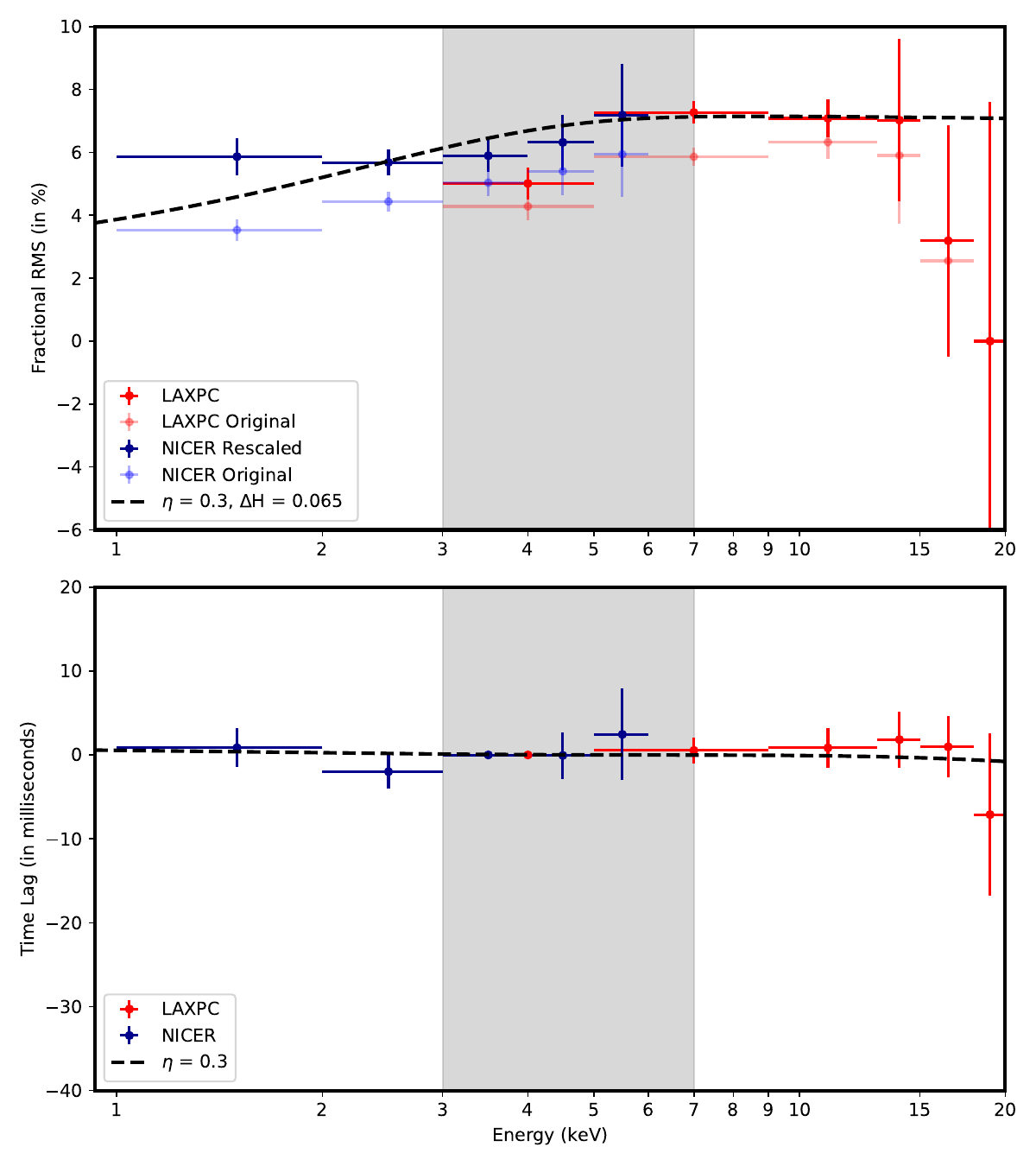}
	\caption{The fractional rms and time lag variation with LAXPC 20 within 3-20 keV and {\em NICER} within 1-7 keV at 9.37 Hz for observation 2. Solid points in the top panel show the re-scaled rms variation and the light points show the original variation that has been obtained from the observation. The solid line represents the model prediction for a corona of size 30 km and $\eta$ = 0.3.}
	\label{Fig8}
\end{figure}

\subsection{LAG and RMS}
For the kHz QPO observed in Obs. 1, we use the LAXPC data to estimate the energy-dependent time lag and r.m.s in the frequency range 625-875 Hz, which are shown in Figure \ref{Fig7}. For time lag, the reference energy band was chosen to be 5-6 keV, since that provided the smallest error bars. The QPO was not detected for energies greater than 9 keV with a high unconstrained upper limit. The r.m.s were found to be nearly constant with a hint of increase in low energies. The  of the time-lags were found to be less than 400 micro-seconds with some evidence for hard lag i.e. the high energy photons are delayed compared to the lower ones.

For the second observation, the simultaneous data from {\em NICER} and LAXPC allows the study of the energy-dependent temporal behaviour over a larger energy range. Figure \ref{Fig8} shows the r.m.s and time-lag variation in the frequency range 4.4-13.4 Hz, covering the broad peak seen in the PDS shown in Figure \ref{Fig6}. The r.m.s increases with energy from ~1 keV to about ~8 keV and saturates after that. In 3-20 keV energy range rms of the QPO comes out to be 11+/-0.2 \%.  While the r.m.s is corrected for the background,  the apparent decrease at high energies may be due to a slight overestimation of the background. The time lag has been computed using the same common reference band of 3-4 keV for both LAXPC and {\em NICER} and is constrained to be less than a few milliseconds. Note that the results from LAXPC and {\em NICER} are consistent with each other.

\subsection{Modeling the RMS and Lag}

The time-lags for kHz QPO are reported to be of the order of $\sim 50 $ microseconds \citep{klitzing:barret2013soft,klitzing:peille2015spectral,klitzing:2018ApJ...860..167T}. Such a short time lag is expected due to delays caused by Compton scattering in a compact source of size $\sim 6$ km \citep{klitzing:lee1998comptonization}. Since the observed time-lags for the
lower Khz QPOs are soft, \citet{lee2001compton} considered a fraction of the Comptonized photons impinging back into the seed photon source. Subsequently, there have been several other works \citep{klitzing:kumar2014energy,klitzing:kumar2016constraining,10.1093/mnras/stz3502,2022MNRAS.515.2099B} which have used this model to the energy-dependent r.m.s and time-lags for Khz QPOs. The model\footnote{\url{http://astrosat-ssc.iucaa.in/data_and_analysis}} predicts the lag and r.m.s behaviour with energy by solving the linearized time-dependent Kompaneets equation  \citep{lee2001compton} in the non-relativistic limit ($kT_e<<m_ec^2$) for small variations in the seed photon temperature or the coronal heating rate. It considers that a fraction, $\eta$, of the Comptonized photons impinges back into the input seed photon source.

The model requires the black body temperature, optical depth and electron temperature of the corona as obtained from the time-averaged spectral fitting. In this work, we have used the XSPEC spectral model {\tt{thcomp*bbody}} which has relativistic corrections, to model the spectra. To be consistent with the timing model, we compared the spectrum from the non-relativistic equation with the best-fit spectrum obtained from {\tt{thcomp*bbody}} and found that for the first observation, the two spectra match if the black body temperature, optical depth and temperature of the corona are chosen to be$\sim 0.7$ keV, $\sim 5.2 $ and $\sim 7$ keV. Thus, for the temporal analysis, we have used these values. Moreover, the observed photon spectrum has a reflection component, which we assume to be not varying. To take this into account, we have re-scaled the r.m.s data points by $C_T/(C_T-C_R)$ where $C_T$ and $C_R$ are the total and reflection count rates for each energy band. Note that the disk emission does not contribute in the LAXPC band i.e. $E> 3$ keV.

The observed time lag for the kHz QPO reported in this work is of the order
of $\sim 200$ micro-seconds and the lag seems to be hard, although the error bars are large. The dashed lines in Figure 6, represent the predicted r.m.s and time lag for a system where the size of the corona is $\sim 30$ km and $\eta \sim 0.1$. A smaller size would result in smaller time lags and larger values of $\eta$ would have predicted soft time lags.

For the second observation, the simultaneous data from {\em NICER} and LAXPC allows us to study the temporal properties of the broadband noise at $\sim 10$ Hz in a wider energy band of 1-20 keV. The best-fit Comptonization model component using {\tt{thcomp*bbody}} matches with the non-relativistic one, if for the latter the black body temperature, the optical depth and corona temperature are chosen to be $\sim 0.72$ keV, $\sim 6.1$ and $\sim 6.7$ keV respectively. We further assume that disk emission and the reflection component are not varying and hence re-scale the observed r.m.s based on the count rate of the components in each energy band.  As shown in the bottom panel of Figure \ref{Fig8}, the time-lags are not detected and are constrained to be less than a few milliseconds. Hence, we use the same size of the corona as for the kHz case i.e. $\sim 30$ km. The dashed lines in Figure \ref{Fig8} show the predicted r.m.s and time-lag for the case when $\eta \sim 0.3$. The significant variability at low energies as observed by  {\em NICER} requires a larger $\eta$ than for the kHz QPO case of Obs. 1.

\section{Summary \& Discussion} \label{sec6}

In this article, we present the broadband spectro-temporal analysis of 4U 1702-429, using {\em AstroSat}  observations at two different epochs separated by a year. The latter observation has a simultaneous {\em NICER} observation, which when combined with {\em AstroSat} data provides both spectrum and rapid timing information in a wide band. LAXPC data reveals that during the second observation, which we call Obs. 2, the source had a higher count rate and was harder than the first. 

The photon spectra of both observations can be described by a thermal Comptonization component whose seed photons arise from a black body spectrum, a disk emission and a non-relativistic reflection component. The size of the black body seed photon source is consistent with being around $\sim 10$ km for both observations for a distance of $\sim 7$ kpc. Thus, we may identify the black body source as the surface of the neutron star. The primary difference between the two observations is that the temperature of the Comptonizing medium (or corona) increases to $\sim 13 $ keV in Obs 2 compared to Obs 1 where it is $\sim 7$ keV, thus making the second observation having a higher flux and harder spectrum than the first. The optical depth of the corona and the seed photon black body temperature are nearly unchanged at $\sim 5$ and $0.7$ keV respectively. An interesting variation is the inner disk radius as inferred from the change in the normalization of the disk component.  For a distance of $\sim 7$ kpc, the inner disk radius decreased from $\sim 150$ km during Obs 1 to $\sim 30$ km in Obs 2. With a fixed iron abundance of 2, \citet{klitzing:2019ApJ...873...99L} estimated the inner disk radius of the source to be $9.0^{+9.6}_{-2.4}$ gravitational radii ($r_g$) using another reflection model {\tt Relxill} \citep{2014ApJ...782...76G}. When the iron abundance was left free, the upper bound of the disk radius increased, resulting in $9.6^{+17.4}_{-0.6} r_g$. For a typical neutron star mass of 1.4 - 2.2 M$_\odot$ , this value of $\sim$ 9.6 $r_g$ corresponds to a physical radius of $\sim$ 45-55 km, with an upper bound of around 130 km. These values are similar to the radius of $124^{+22}_{-18}$ in softer state and $32.8^{+6.4}_{-15.6}$ km in relatively harder state obtained in our analysis. We note that the analysis undertaken by \citet{klitzing:2019ApJ...873...99L} was using {\em NuSTAR} data which is sensitive above 3 keV and hence their model would not constrain the disc component effectively. We note that if instead of {\tt ireflect} we use {\tt Relxill} to model the reflection component, we get a significantly higher $\chi_{red}^2$ (i.e. 139.37/104 and 253.23/204 for Obs 1 and 2) and the inner disc radii are not constrained. This is most probably because {\tt Relxill} assumes that the  incident emission is due to thermal Comptonization of a black body at a very low temperature of 0.1 keV, while the convolution model {\tt ireflect} allows for the incident emission to be the one used to fit the low energy part of the spectrum. The reflection fraction does not change between the observations, implying that the component is perhaps from a distant reflector.

The spectral modeling reported in this work differs from that undertaken by \citet{2024MNRAS.529.2234V} for the persistent emission, by considering the hard X-ray emission to be from thermal Comptonization instead of a power-law. The spectral parameters obtained for the hard state (i.e. Obs 2 of this analysis, Table \ref{Table2}) are similar to the values reported  by \citet{Banerjee_2024}. They report a coronal temperature of $>$ 3 keV with a  black body temperature and radius of $\sim$ 0.5 keV and $\sim$ 40 km. We note that for the {\em AstroSat} observation analysis, they fix the coronal temperature to be at 100 keV, while in our analysis we find the value to be $\sim$ 13 keV which is similar to what they used for {\em NICER} analysis. For the soft state (i.e. Obs 1 of this analysis, Table \ref{Table2}), they report a significantly higher value of the black body temperature of $\sim 1.8$ keV (for the closest {\em NICER} observation of  March 2018) as compared to $\sim 0.7$ keV reported here. Their inner disk temperature is also higher with $\sim 0.7$ keV compared to $\sim 0.4$ keV reported here and find  a smaller disc radius of $\sim 10$ kms. We attribute these differences for the soft state as being due to the wider energy band of {\em AstroSat} used in this work.

Timing analysis reveals a kHz QPO at $\sim 800$ Hz for Obs 1. Energy-dependent time-lags at the kHz QPO are constrained to be less than a few 100 microseconds. We show that the energy-dependent fractional rms and time-lags are consistent with the interpretation that there is variation in the heating rate of the corona and a fraction of the Comptonized photons impinge onto the seed photon source. Here the time-lags are interpreted as being due to Compton scattering and the size of the corona can be constrained to be $\leq 30$ km. For Obs 2., where there is simultaneous {\em NICER} and LAXPC data, no QPOs are observed, and instead there is broadband noise around $\sim 10$ Hz. For this feature, the time-lags are constrained to be less than a few milliseconds and a similar model as used for the kHz QPO can explain the energy-dependent r.m.s and time-lags.

The results presented in this work show that for 4U 1702-429, the primary X-ray emission is from a Comptonizing medium surrounding a black body source which can be identified as the surface of the neutron star.  For Obs 1, the inner disk radius is rather large at $\sim 150$ km, while the size of the Comptonizing medium is constrained to be less than $30$ km based on the observed time-lags. This implies that there is a region between the disk and the neutron star which is probably a radiatively inefficient flow. It is during this observation that the source shows a kHz QPO at $\sim 800$ Hz, which must be associated with a radius close to the neutron star. Hence the kHz QPO cannot be associated with the inner disk radius, but instead perhaps is related to the extent of the Comptonizing medium. On the other hand, Obs 2, reveals a hotter corona, an inner disk radius closer to the neutron star and no kHz QPO.

The results are based on two observations of 4U 1702-429, and clearly, there is a need for a larger number of observations covering the different spectral states of the source using wide-band spectroscopy from simultaneous {\em NICER} and {\em AstroSat} observations. Such a study will confirm whether a large truncated disk radius is a necessary condition for the occurrence of kHz QPOs and whether a closer disk radius inhibits the occurrence. Moreover, such observations with longer exposures than the data analyzed in this work will provide more stringent constraints on energy-dependent r.m.s and time-lags over a wide energy band, which may then be used to further constrain the geometry and parameters of the Comptonizing corona.

\section{acknowledgments}

We have utilized the data from LAXPC and SXT payloads onboard {\it  AstroSat} available at ISSDC and the data of {\em NICER} XTI available on the HEASARC website. We are thankful to ASSC (AstroSat Science Support Cell ) for providing their support from time to time by solving the issues regarding scientific analysis. Furthermore, SC is also thankful to IUCAA for providing the periodic visit to carry out most of the work presented here. We wish to further acknowledge the LAXPC Payload Operation Center (POC) and SXT POC at TIFR, Mumbai, and the calibration team of the {\em NICER}. SM and SKP would also like to thank IUCAA for the visiting associateship. The research leading to these results has been funded by the Department of Space, Govt. of India, ISRO under grant no. DS\_2B-13013(2)/10/2020-Sec.2. 

\section{Data Availability}
The LAXPC  and SXT archival data that has been used in this article can be found at {\it AstroSat} ISSDC website (\url{ https://astrobrowse.issdc.gov.in/astro\_archive/archive}). The {\em NICER} data are publicly available at the website (\url{https://heasarc.gsfc.nasa.gov/cgi-bin/W3Browse/w3browse.pl}).


\bibliography{ms}{}
\bibliographystyle{aasjournal}



\end{document}